\documentstyle[12pt]{article}
\begin{document}
\noindent
\begin{center}
{\Large {\bf Comments on Scalar-Tensor \\Representation of $f(R)$
 Theories\\}} \vspace{2cm}
 ${\bf Yousef~Bisabr}$\footnote{e-mail:~y-bisabr@srttu.edu.}\\
\vspace{.5cm} {\small{Department of Physics, Shahid Rajaee Teacher
Training University,
Lavizan, Tehran 16788, Iran.}}\\
\end{center}
\vspace{1cm}
\begin{abstract}
We propose a scalar-tensor representation of $f(R)$ theories with
use of conformal transformations.  In this representation, the
model takes the form of the Brans-Dicke model with a potential
function and a non-zero kinetic term for the scalar field.  In
this case, the scalar field may interact with matter systems and
the corresponding matter stress tensor may be non-conserved.

\end{abstract}
~~~~~~~PACS Numbers: 98.80.-k \vspace{3cm}
\section{Introduction}
Recent observations on expansion history of the universe indicate
that the universe is experiencing a phase of accelerated expansion
\cite{sup}.  This can be interpreted as evidence either for
existence of some exotic matter components or for modification of
the gravitational theory.  In the first route of interpretation
one can take a perfect fluid with a sufficiently negative
pressure, dubbed dark energy \cite{mel}, to produce the observed
acceleration. There is also a large class of scalar field models
in the literature including, quintessence \cite{q}, phantom
\cite{ph} and quintom fields \cite{qui} and so forth. In the
second route, however, one attributes the accelerating expansion
to a modification of general relativity. A particular class of
models that has recently drawn a significant amount of attention
is the so-called $f(R)$ gravity models \cite{car}\cite{r}. These
models propose a modification of Einstein-Hilbert action so that
the scalar curvature is replaced by some arbitrary function
$f(R)$. It is well known that $f(R)$ theories of gravity can be
written as a scalar-tensor theory by applying a Legendre
transformation \cite{soti}\cite{soko}.  This scalar-tensor
representation corresponds to a class of Brans-Dicke theory with a
potential function and $\omega=0$ in the metric formalism. There
is also such a correspondence for $\omega=-\frac{3}{2}$ in the
Palatini formalism in which metric and connections are taken as
independent variables, see \cite{sot} and references therein. Here
we do not consider Palatini formalism.\\ Although $f(R)$ gravity
theories exhibit a natural mechanism for accelerated expansion
without recourse to some exotic matter components, because of
vanishing of the kinetic term of the scalar field in the
scalar-tensor representation there are criticisms that emphasized
inability of these models to pass solar system tests \cite{chiba}.
In the present note we will focus on the dynamical equivalence of
$f(R)$ theories and the Brans-Dicke theory with use of conformal
transformations. We will show that this equivalence holds for an
arbitrary Brans-Dicke parameter.  In this case, however, the
gravitational coupling of matter systems may be anomalous in
the sense that the scalar field inter the matter field action. \\
~~~~~~~~~~~~~~~~~~~~~~~~~~~~~~~~~~~~~~~~~~~~~~~~~~~~~~~~~~~~~~~~~~~~
\section{The Model}
To begin with, we offer a short review on the equivalence of
$f(R)$ theories with a particular class of Brans-Dicke theory with
a potential. We consider the following action\footnote{We work in
the unit system in which $\hbar=c=8\pi G=1$.}
\begin{equation}
S=\frac{1}{2}\int d^4 x \sqrt{-g} f(R) +S_{m}(g_{\mu\nu}, \psi)
\label{a1}\end{equation} where $f(R)$ is an arbitrary function of
the scalar curvature $R$.  The matter action $S_{m}(g_{\mu\nu},
\psi)$ is
\begin{equation}
S_{m}(g_{\mu\nu}, \psi)=\int d^4 x \sqrt{-g}~L_{m}(g_{\mu\nu},
\psi)
\end{equation}
 in which the Lagrangian density $L_{m}$ corresponds to matter fields which are
collectively denoted by $\psi$. One usually introduces a new field
$\chi=R$ by which the action (\ref{a1}) can then be written as
\begin{equation}
S=\frac{1}{2}\int d^4 x \sqrt{-g}~ \{f(\chi)+f^{'}(\chi)(R-\chi)\}
+S_{m}(g_{\mu\nu}, \psi) \label{a2}\end{equation} where prime
denotes differentiation with respect to $R$. Now variation with
respect to $\chi$ leads to the equation
\begin{equation}
f^{''}(R)(\chi-R)=0 \end{equation} If $f^{''}(R)\neq 0$, we have
the result $\chi=R$.  Inserting this result into (\ref{a2})
reproduces the action (\ref{a1}). Then redefining the field $\chi$
by $\Phi=f^{'}(R)$ and setting
\begin{equation}
V(\Phi)=\chi(\Phi)\Phi-f(\chi(\Phi)) \label{a3}\end{equation} the
action (\ref{a2}) takes the form
\begin{equation}
S=\frac{1}{2}\int d^4 x \sqrt{-g}~ \{\Phi R-V(\Phi)\}
+S_{m}(g_{\mu\nu}, \psi) \label{a4}\end{equation} This is the
Brans-Dicke action with a potential $V(\Phi)$ and a Brans-Dicke
parameter $\omega=0$.  Therefore there is a dynamical equivalence
between $f(R)$ theories and a class of Brans-Dicke theories with a
potential function.  The important point in the above
transformation is that the matter sector is remained unchanged. In
particular, in this representation of $f(R)$ theories the matter
action $S_{m}(g_{\mu\nu}, \psi)$ is independent of the scalar
field $\Phi$.  Thus in both actions (\ref{a1}) and (\ref{a4}) the
weak equivalence principle holds and test particles follow
geodesics
lines of $g_{\mu\nu}$.\\
In original form of the Brans-Dicke theory \cite {bd}, where the
potential term of the scalar field is not present, it is found
that in order to get agreement between predictions and solar
system experiments $\omega$ should be large and positive. The
current observations set a lower bound on $\omega$ which is
$\omega > 3500$ \cite{will}. If the theory is allowed to have a
potential, the scalar field should be very light and should
mediate a gravity force of long range which is not consistent with
solar system experiments \cite{chiba}. On the other hand, it is
shown \cite{fara} that the dynamical equivalence of (\ref{a4}) and
$f(R)$ theories is ill-defined in the scale of solar system. The
underlying logic is based on the fact that there is no observed
deviation from general relativity at this scale and $f(R)$
theories must be reduced to general relativity in an appropriate
limit. The problem is that general relativity corresponds to
$f(R)=R$ for which $f^{''}(R)=0$, while the
dynamical equivalence requires $f^{''}(R)\neq 0$.\\
We intend here to use a different scalar-tensor representation of
$f(R)$ theories suggested by conformal transformations.  To this
aim, we apply the following conformal transformation
\begin{equation}
\tilde{g}_{\mu\nu}=\Omega_{1}~g_{\mu\nu}~,~~~~~~~~~~~~~~~~~~~~~~~~\Omega_{1}=f^{'}(R)
\label{a5}\end{equation} to the action (\ref{a1}).  This together
with a redefinition of the conformal factor in terms of a scalar
field $\phi=\sqrt{\frac{3}{2}}\ln \Omega_{1}$, yields \cite{soko}
\begin{equation}
S=\frac{1}{2}\int d^4 x ~\sqrt{-\tilde{g}}~ \{
\tilde{R}-\tilde{g}^{\mu\nu}\nabla_{\mu}\phi\nabla_{\nu}\phi-V(\phi)\}
+ S_{m}(\tilde{g}_{\mu\nu}, \psi, \phi)
 \label{a6}\end{equation}
where
\begin{equation}
S_{m}(\tilde{g}_{\mu\nu}, \psi, \phi)=\int d^4 x
\sqrt{-\tilde{g}}~e^{-2\sqrt{\frac{2}{3}}\phi}~L_{m}(\tilde{g}_{\mu\nu},
\psi) \label{a8}\end{equation} In the action (\ref{a6}), $\phi$ is
minimally coupled to $\tilde{g}_{\mu\nu}$ and appear as a massive
self-interacting scalar field with a potential
\begin{equation}
V(\phi)=\frac{1}{2}e^{-\sqrt{\frac{2}{3}}\phi}r[\Omega_{1}(\phi)]-\frac{1}{2}e^{-2\sqrt{\frac{2}{3}}\phi}
f(r[\Omega_{1}(\phi)]) \label{a7}\end{equation} where the function
$r(\Omega_{1})$ is the solution of the equation
$f^{'}[r(\Omega_{1})]-\Omega_{1}=0$. Thus the variables
$(\tilde{g}_{\mu\nu}, \phi)$ in the action (\ref{a6})
provide the Einstein frame variables for $f(R)$ theories.\\
Variation of (\ref{a6}) with respect to $\tilde{g}_{\mu\nu}$ and
$\phi$ give, respectively,
\begin{equation}
\tilde{G}_{\mu\nu}=t_{\mu\nu}+e^{-\sqrt{\frac{2}{3}}\phi}T_{\mu\nu}(\tilde{g}_{\mu\nu},
\psi)
\end{equation}
\begin{equation}
\tilde{\Box}\phi-\frac{dV(\phi)}{d\phi}=\sqrt{\frac{1}{6}}e^{-2\sqrt{\frac{2}{3}}\phi}T
\end{equation}
where
\begin{equation}
T_{\mu\nu}=\frac{-2}{\sqrt{-g}}\frac{\delta S_{m}}{\delta
g^{\mu\nu}}
\end{equation}
\begin{equation}
t_{\mu\nu}=\partial_{\mu}\phi
\partial_{\nu}\phi-\frac{1}{2}\tilde{g}_{\mu\nu}\partial_{\gamma}\phi
\partial_{\gamma}\phi-V(\phi)\tilde{g}_{\mu\nu}
\end{equation}
and $T \equiv g^{\mu\nu}T_{\mu\nu}$.  Now the Bianchi identity
implies \begin{equation}
\tilde{\nabla}^{\mu}T_{\mu\nu}=\sqrt{\frac{2}{3}}\partial^{\mu}\phi~
T_{\mu\nu} -\sqrt{\frac{1}{6}}
e^{-\sqrt{\frac{2}{3}}\phi}\partial_{\nu}\phi ~T
\end{equation}
There is also a scalar-tensor representation by applying the
conformal transformation
\begin{equation}
\tilde{g}_{\mu\nu}=\Omega_{2}~
\bar{g}_{\mu\nu}~,~~~~~~~~~~~~~~~~~~~~~~\Omega_{2}=e^{\frac{\phi}{\alpha}}
\label{a9}\end{equation} to the action (\ref{a6}) with $\alpha$
being a constant parameter. This together with $\phi=\alpha \ln
\varphi$ transform the action (\ref{a6}) to
\begin{equation}
S=\frac{1}{2}\int d^4 x~ \sqrt{-\bar{g}}~ \{\varphi
\bar{R}-\frac{\omega}{\varphi}~\bar{g}^{\mu\nu}~\nabla_{\mu}\varphi\nabla_{\nu}\varphi-U(\varphi)\}
+ S^{'}_{m}(\bar{g}_{\mu\nu}, \psi, \varphi)\label{a10}
\end{equation} where
\begin{equation}S^{'}_{m}(\bar{g}_{\mu\nu}, \psi,
\varphi)=\int d^{4}x ~\sqrt{-\bar{g}}~\varphi^{n}
~L_{m}(\bar{g}_{\mu\nu}, \psi, \varphi) \label{a11}\end{equation}
and
\begin{equation}
\omega = \alpha^{2}-\frac{9}{2}\label{a12} \end{equation}
\begin{equation}
n = 2-2\alpha \sqrt{\frac{2}{3}} \label{a13}\end{equation}
\begin{equation}
U(\varphi)=2\varphi^{2} V(\varphi) \label{a14}\end{equation} This
is the scalar-tensor representation of the action (\ref{a1})
obtaining by conformal transformations (\ref{a5}) and (\ref{a9}).
In contrast with this representation, the gravitational part of
the action (\ref{a1}) consists only of the metric tensor
$g_{\mu\nu}$ which obeys fourth-order field equations. We may call
the conformal frames corresponding to the actions (\ref{a1}) and
(\ref{a10}) the Jordan frame representation of
(\ref{a6})\footnote{ Note that we define here Jordan frame in
terms of how the geometry is described in the vacuum sector
        rather than in terms of how it couples with matter systems \cite{soko}. The action (\ref{a1}) is in Jordan
        frame since the resulting field equations are fourth-order in terms of metric tensor.
        On the other hand, the action (\ref{a10}) is also in Jordan frame since it describes the geometry
        by a metric tensor and a scalar field (nonminimal coupling of the scalar field).}.  Here a
question which arises is that which of the conformal frames
corresponding to the actions (\ref{a1}), (\ref{a6}) and
(\ref{a10}) should be considered as the physical frame. It should
be pointed out that reformulation of a theory in a new conformal
frame leads, in general, to a different physically inequivalent
theory.  The ambiguity of the choice of a particular frame as the
physical one is a longstanding problem in the context of conformal
transformations.  The term ``physical" theory denotes one that is
theoretically consistent and predicts values of some observables
that can, at least in principle, be measured in experiments
\cite{fgn}. In this respect different authors may consider
different conformal frames as physical according to their attitude
towards the issue of the conformal frames\footnote{For a good
review on this issue, see \cite{fgn} and references therein.}. For
instance, while in $f(R)$-theories one usually takes the Jordan
frame as the physical frame one may consider positivity of energy
and stability to consider the Einstein conformal frame as the
physical one \cite{soko}.  Thus the choice of a physical frame
between the representations (\ref{a1}), (\ref{a6}) and (\ref{a10})
should
be based on the physical outcomes of the corresponding models.\\
Let us now compare the two scalar-tensor representation of $f(R)$
theories, namely the actions (\ref{a4}) and (\ref{a10}).  There
are two important differences: Firstly, in the action (\ref{a10})
the Brans-Dicke scalar field $\varphi$ has a non-zero kinetic term
and the Brans-Dicke parameter $\omega$ is only constrained by
observations.  Secondly, the scalar field $\varphi$ enter the
matter part of the action (\ref{a10}).  The latter means that the
scalar field interacts with matter systems and tests particles do
not follow the geodesic lines of the metric $\bar{g}_{\mu\nu}$. We
will return to this issue later.\\
Variation with respect to $\bar{g}_{\mu\nu}$ and $\varphi$ leads
to the field equations
\begin{equation} \varphi
\bar{G}_{\mu\nu}-\frac{\omega}{\varphi}(\nabla_{\mu}\varphi
\nabla_{\nu}\varphi-\frac{1}{2}\bar{g}_{\mu\nu}\nabla_{\gamma}\varphi
\nabla^{\gamma}\varphi)-(\bar{\nabla}_{\mu}\nabla_{\nu}\varphi-\bar{g}_{\mu\nu}\bar{\Box}\varphi)
+\frac{1}{2}U(\varphi)\bar{g}_{\mu\nu}=\bar{T}_{\mu\nu}
\label{a15}\end{equation}
\begin{equation}
\frac{2\omega}{\varphi}\bar{\Box}\varphi-\frac{\omega}{\varphi^{2}}\nabla_{\gamma}\varphi
\nabla^{\gamma}\varphi+\bar{R}-\frac{dU(\varphi)}{d\varphi}=\varphi^{-1}\bar{T}
\label{a16}\end{equation} where
\begin{equation}
\bar{T}_{\mu\nu}=-\frac{2}{\sqrt{-\bar{g}}}\frac{\delta
S_{m}}{\delta \bar{g}^{\mu\nu}} \label{16}\end{equation} and
$\bar{T}=\bar{g}^{\mu\nu}\bar{T}_{\mu\nu}$ is the trace of the
stress tensor $\bar{T}_{\mu\nu}$.  Now applying the Bianchi
identity $\bar{\nabla}^{\mu}\bar{G}_{\mu\nu}=0$ and using the
field equation of the scalar field (\ref{a16}), we obtain
\begin{equation}
\bar{\nabla}^{\mu}\bar{T}_{\mu\nu}=-a_{\nu}
\label{a17}\end{equation} where
\begin{equation}
a_{\nu}=\frac{1}{2}~\bar{T}~\partial_{\nu}\ln\varphi
\label{a18}\end{equation} The equation (\ref{a17}) implies that
the matter stress tensor $\bar{T}_{\mu\nu}$ is not conserved due
to interaction of the scalar field $\varphi$ with the matter part
of (\ref{a10}). Except for the case that the matter field action
(\ref{a11}) is traceless \cite{soko}, the scalar field $\varphi$
influences the motion of any gravitating matter. In fact,
$a_{\nu}$ indicates an anomalous acceleration corresponding to a
fifth force. \\It should be noted that there are different types
of models in literature that concern matter systems that are not
conserved due to interaction with an arbitrary function of scalar
curvature \cite{geo} or some scalar fields \cite{sca}. However,
the important difference between (\ref{a17}) and the corresponding
equations in those models is that the former is the result of the
well-known property of conformal transformations, namely that
conservation equation of a matter stress
tensor with a nonvanishing trace is not conformally invariant \cite{wald}. \\
It is possible to apply this result in the scale of solar system.
To do this, we first note that combining (\ref{a5}) and (\ref{a9})
gives the relation between the scalar field $\varphi$ and the
function $f(R)$
\begin{equation}
\varphi=[f^{'}(R)]^{\frac{2}{2-n}} \label{a19}\end{equation} Then
we take $\bar{T}_{\mu\nu}$ to be the stress tensor of dust (or
perfect fluid with zero pressure) with energy density
$\bar{\rho}$.  In this case and for a static spacetime we obtain
for the spatial part of $a_{\nu}$,
\begin{equation}
a_{i}=\frac{1}{2-n}~\bar{\rho}~\partial_{i}R~\frac{f^{''}(R)}{f^{'}(R)}
\label{a20}\end{equation} As one expects, in the case that $f(R)$
is a linear function of $R$ like the Eistein-Hilbert action, the
anomalous acceleration is zero. For CDTT model \cite{car} in which
$f(R)=R-\frac{\mu^{4}}{R}$  we obtain
\begin{equation}
a_{i}=\frac{2}{n-2}~\bar{\rho}~\partial_{i}\ln
R~(1+\frac{R^2}{\mu^4})^{-1} \label{a21}\end{equation} where $\mu$
is an arbitrary mass scale.
~~~~~~~~~~~~~~~~~~~~~~~~~~~~~~~~~~~~~~~~~~~~~~~~~~~~~~
\section{Concluding Remarks}
While the scalar-tensor representation (\ref{a4}) of $f(R)$
theories is useful in cosmological scales it suffers problems in
weak field limit and solar system scales. Firstly, the kinetic
term of the scalar field vanishes which is in conflict with
current bounds on the value of $\omega$. Secondly, it is recently
reported that since $f^{''}(R)=0$ in this scales, the dynamical
equivalence of $f(R)$ theories and scalar-tensor theories
represented by (\ref{a4}) breaks down.  The main feature of our
analysis is to show that there is also a dynamical equivalence
between $f(R)$ theories and scalar-tensor theories with use of
conformal transformations.  In this representation the scalar
field has a non-vanishing kinetic term and a non-zero Brans-Dicke
parameter. Therefore, the observational constraints can be applied
on $\omega$ in this representation.  However, it should be noted
that the action (\ref{a10}) differs from the Brans-Dicke action in
two ways. Firstly, it contains a potential function $U(\varphi)$.
Secondly, the matter system interacts with the scalar field
$\varphi$. That the scalar field possess a potential function
clearly alters the usual bound on the Brans-Dicke parameter
$\omega$ so that the new bound depends on the functional form of
the potential \cite{ber}.  On the other hand, the coupling of the
scalar field with matter sector should be strongly suppressed so
as not to lead to observable effects. In fact, one can use
chameleon mechanism \cite{welt} to implement constraints on the
potential function $U(\varphi)$.  Then using a relation between
$\varphi$ and the curvature scalar (see the relation (\ref{a19})),
this can provide some viable forms of $f(R)$ theories which are in
accord with local gravity tests.  Indeed, this is the method that
is recently used by some authors to deal with $f(R)$ theories
which are consistent with Solar System experiments \cite{cham}. \\
It is important to note that non-conservation of the stress tensor
$\bar{T}_{\mu\nu}$ should not be considered as an intrinsic
behavior of the model presented here.  It is simply related to the
fact that before applying the conformal transformations, the
matter action is introduced in the nonlinear action (\ref{a1}).
The matter action might be added after the conformal
transformations in the Jordan frame.  In that case there would be
no anomalous acceleration. This ambiguity of introducing matter
systems to equivalent conformal frames is closely related to the
well-known problem that which of these frames should be taken as
the physical one \cite{soko}\cite{fgn}. Without dealing with this
long-standing problem, we would like to point out that the
advantage of the non-minimal coupling of matter in (\ref{a10}) is
that it can potentially explain the possible deviations of $f(R)$
theories (or its scalar-tensor representation) from newtonian
gravity in local experiments. Moreover, as earlier stated, the
scalar field may also be interpreted as a chameleon field which
can suppress detectable effects of anomalous gravitational
coupling of matter in solar system scales. This chameleon behavior
of the scalar field is under progress by the author and is
deserved to be investigated elsewhere. 
%%%%%%%%%%%%%%%%%%%%%%%%%%%%%%%%%%%%%%%%%%%%%%%%%%%%%%%%%%%%%%%%%%%%%%%%%%%%%%

\end{document}